\documentclass{aa}
\usepackage{graphicx}
\usepackage{color}

\usepackage{graphicx,natbib}

\bibpunct{(}{)}{;}{a}{}{,}

\def\apj{ApJ}

\def\aap{A\&A}

\begin{document}

\sloppypar

   \title{Diagnostics of the black hole candidate SS433 with the RXTE}

   \author{Filippova E.\inst{1,2}, Revnivtsev M.\inst{1,2}, Fabrika
S.\inst{3},  Postnov K.\inst{4}, Seifina E.\inst{4}}

   \offprints{kate@hea.iki.rssi.ru}

   \institute{ 
              Max-Planck-Institute f\"ur Astrophysik,
              Karl-Schwarzschild-Str. 1, D-85740 Garching bei M\"unchen,
              Germany 
     \and
              Space Research Institute, Russian Academy of Sciences,
            Profsoyuznaya 84/32, 117997 Moscow, Russia  
      \and
              Special Astrophysical Observatory, Nizhnij Arkhyz,
Karachaevo-Cherkesiya, 369167,  Russia
         \and
              Sternberg Astronomical Institute, 119992, Moscow, Russia
            }
  \date{}

        \authorrunning{Filippova et al.}

   \abstract{We present our analysis of the extensive monitoring of SS433
by the RXTE observatory collected over the period 1996-2005. The
difference between energy spectra taken at different precessional
and orbital phases shows the presence of strong photoabsorption
($N_{\rm H}>10^{23}$\,cm$^{-2}$) near the optical star, 
probably due to its powerful, dense wind.
Therefore the size of the secondary deduced from analysis of
X-ray orbital eclipses might be significantly larger than its Roche
lobe size, which must be taken into account when
evaluating the mass ratio from analysis of X-ray eclipses.
Assuming that a precessing accretion disk 
is geometrically thick, we recover the temperature
profile in the X-ray emitting jet that best fits the observed
precessional variations in the X-ray emission temperature.
The hottest visible part of the X-ray jet is located at
a distance of $l_0/a\sim0.06-0.09$, or $\sim2-3\times10^{11}$cm from the
central compact object, and has a temperature of about $T_{\rm max}\sim30$ keV. 
We discovered appreciable orbital X-ray
eclipses at the ``crossover'' precessional phases (jets are in the
plane of the sky, disk is edge-on), which under model assumptions
put a lower limit
on the size of the optical component $R/a\ga0.5$ and an upper
limit on a mass ratio of binary companions $q=M_{\rm x}/M_{\rm
opt}\la0.3-0.35$, if the X-ray opaque size of the star is not
larger than $1.2R_{\rm Roche, secondary}$. 
   \keywords{accretion, accretion disks--
                black hole physics --
                instabilities --
                stars:binaries:general --
                X-rays: general  --
                X-rays: stars
               } 
   }

   \maketitle

%

\section{Introduction}

The source SS433 is the only Galactic X-ray binary with X-ray emission
from optically thin thermal plasma originating in hot jets outflowing with a sub-relativistic
velocity of 0.26 $c$.
The binary system is thought to consist of a compact object
(probably a black hole) accreting matter at a super-Eddington rate from a high-mass star
filling its Roche lobe (\citealt{margon84},
\citealt{cherep02}, see \citealt{fabrika04} for a recent review).
Apparently, the innermost accretion flow and the supercritical accretion disk wind
completely screen the region
of the main energy release, and most of the energy from SS433
is observed in the optical and UV
(Cherepashchuk et al., 1982; Dolan et al., 1997).

The system demonstrates a complex variability including periodicities -
the precessional ($\sim$162 days), orbital
($\sim$13 days), and nutational (nodding,$\sim 6$ days) periods - which
can be used to tightly constrain the binary system parameters. In
particular, the kinematical model of the system leads to a
binary inclination angle  of $i\sim 78.05^\circ$, and the jet
precession angle is $\sim 20.92^\circ$ \cite{fabrika04}.

Mildly relativistic ($v\sim0.26$c) jets launched in the vicinity of the compact object
consist of protons and electrons (``heavy'' jet) with a high temperature
(1-30 keV). Plasma moving along the jet gradually cools down so that,
at distances $\sim 10^{13}-10^{14}$ cm from the central object, a
thermal instability develops and clumps start forming.
At distances $10^{14}-10^{15}$ cm, the temperature of matter
drops to $\sim2\times10^{4}$K so that optical line emission appears.
At larger distances the jets shine in the radio diapason.

The main properties of the X-ray emission observed from SS433
can be summarized as follows:

\begin{itemize}

\item The X-ray emission mechanism is thermal bremsstrahlung radiation
\citep[e.g.][]{marshall79,w96,matsuoka86,brinkmann91}. 
Based on measurements of the doppler shifts of the emission lines
from the thermal plasma, it is believed that the total (or majority of) 
the X-ray emission originates in the jets \citep{fabrika04}.

\item The temperature of optically thin plasma decreases along the jet.
The hottest regions near the jet base have temperatures up to $\sim$20-30 keV
\citep[e.g.][]{kotani96,marshall02,cherep03,namiki03}.

\item A detailed study of X-ray emission lines in SS433 shows that
the full opening angle of the jet should be very small, not larger than $\sim
1-2^\circ$. Such an opening angle roughly corresponds to free expansion
of the jet material in the direction perpendicular to the jet motion
\citep{marshall02,namiki03}.

\item
There are X-ray eclipses caused by the donor star. The temperature
of the observed radiation drops significantly during the eclipse, and
the depth of the eclipse increases at high energies
\citep{stewart87,kawai89,brinkmann91}.

\item Precession of the geometrically thick accretion disk is accompanied
by strong variations in both temperature and the flux of X-ray
emission \citep{yuan95}. Amplitudes of precessional and orbital variations in
X-ray emission strongly increase with energy
\citep[e.g.][]{cherep05}. The precession variations
are likely to be due to partial eclipse of the X-ray jets by
the thick precessing accretion disk.

\end{itemize}

Over the life time of the RXTE observatory many
observations of SS433 were performed at different precessional and orbital phases
(see e.g.\citealt{gies02a,nandi05}). The relatively complete
sampling of the source at different orbital and precessional phases
provides us with an opportunity to perform a tomographic study of the
X-ray jet in SS433. Examining of orbital eclipses 
(caused by the optical star) and
precessional variability (caused by the geometrically thick accretion disk)
allows us to separate contributions of different
parts of the jet to the total X-ray emission.

\section{Observations and data analysis}

In this paper, we have used all publicly available data of
the RXTE observatory obtained from April 1996 until August 2004 and
also the results of our dedicated SS433 observations during RXTE
AO10 (P91103). In total, this includes 100 observations taken at
different precessional and orbital phases of the system. We 
used precessional and orbital ephemeris furnished by
\cite{fabrika04}.  We adopted the following parameters of the system:
the moment of the
maximum emission-line separation (T3) $T_3=2443507.47$ JD, the
precessional period $P_{\rm prec}=162.375$ days, the orbital period
$P_{\rm orb}=13.08211$ days, the moment of the primary optical
eclipse $T_0=2450023.62$ JD. For nutational periodicity, we
adopted $P_{\rm nut}=6.2877$ days, the moment of maximum brightness
in V band due to  the nodding variability $T_{\rm 0,nut}=2450000.94$ JD,
and the nodding amplitude $2.8^\circ$.

Standard tasks of the LHEASOFT/FTOOLS
5.3.1 package were utilized for data processing.
For accurate background modeling of the PCA
spectrometer of the RXTE observatory, we applied the ``L7\_240CM'' faint
model \cite[see e.g.]{jahoda06}. Only spectra obtained by the PCU2
detector were used for the analysis.
To construct broad-band spectra, we also used data from HEXTE detectors.

\section{Variation in X-ray emission over precessional and orbital periods}

\begin{figure}[htb]
\begin{center}
\includegraphics[width=0.7\columnwidth]{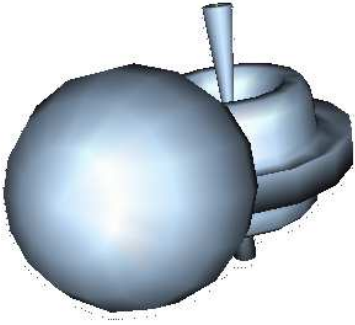}
\end{center}
\caption{Scheme of the X-ray binary SS433. Both the optical component and
geometrically-thick accretion disk can eclipse central
parts of the X-ray jet.} \label{scheme}
\end{figure}

The binary system is schematically shown in Fig.\ref{scheme}.
The binary separation is assumed to be $a\sim 4\times
10^{12}$ cm  \citep[e.g.][]{hillwig04}. The mass ratio of the
companions  is still not well-constrained and varies in different
works from $q=M_{\rm x}/M_{\rm opt}\sim 0.2$ to $q\sim 0.6$
\citep{antokhina92,gies02}. However,
the latest studies of the
kinematics of the binary system favor a smaller mass ratio $q\sim0.2-0.3$ 
\citep{hillwig04,cherep05}.

 After discovery of the donor-star absorption spectrum 
in the blue spectral region of SS433 \citep{gies02},
there were several contradictory results for the orbital behavior
of the absorption lines. From our experience  with 
spectral studies of these absorption lines \citep{cherep05}, we may conclude that   
there are two important restrictions, hand by adhering to them one may hope to detect 
"real donor-star" absorption lines. 

1. Observations must be performed in
the precessional phases of the most open accretion disk ($\psi \sim 0$), 
when the disk outflow does not intersect the line of sight. The observations 
by \cite{gies02}, \cite{hillwig04} and \cite{cherep05} satisfy this condition.

2. Only the weakest absorption lines have to be taken into account, while
stronger lines will trace the gas streams in the binary even at precessional phases 
$\psi \sim 0$, because of the huge ($\sim 10^{-4}M_{\odot}$/yr) mass transfer 
rate in SS433 (\citealt{fabrika04}). In the commonly used cross-correlation 
method, the strongest Fe\,II absorption lines in the studied spectral region 
will dominate and distort the radial velocity curves \citep{hillwig04,barnesetal06}. 
The observations by \cite{gies02} and \cite{cherep05} satisfy the second restriction.  
Great care must be taken when deriving and interpreting the behavior 
of these absorption 
features \citep{barnesetal06}. The heating of the donor surface revealed by
\cite{cherep05} makes the task even more complicated.

In SS433, the optical star loses matter both via 
the Roche lobe overflow and a strong stellar wind.
Matter that goes through  the inner Lagrangian point forms an
accretion disk around the compact object, which eventually becomes
geometrically thick because of strong super-Eddington accretion (see e.g. \citealt{abr04} and 
references therein).
Both simple estimates and more detailed calculations show that this
geometrically thick accretion disk should have a height comparable
to its radius $H/R\sim 1$ \citep[e.g.][]{okuda05,nazarenko05}.

 The observational appearance of SS433 in X-rays is determined by
multi-temperature optically thin thermal plasma emission of the jet. The X-ray emission
from SS433 is subjected to both systematic and chaotic
variations. Chaotic variations are
likely to be caused by self-similar variability in the instantaneous mass
accretion rate in the disk \citep{mikej05}. Systematic variations in
 X-ray luminosity and spectral shape are determined by
precessional (phases $\psi$) and orbital (phases $\phi$) motions in
the binary system.

\begin{figure}[htb]
\includegraphics[width=\columnwidth,bb=33 186 570 654,clip]{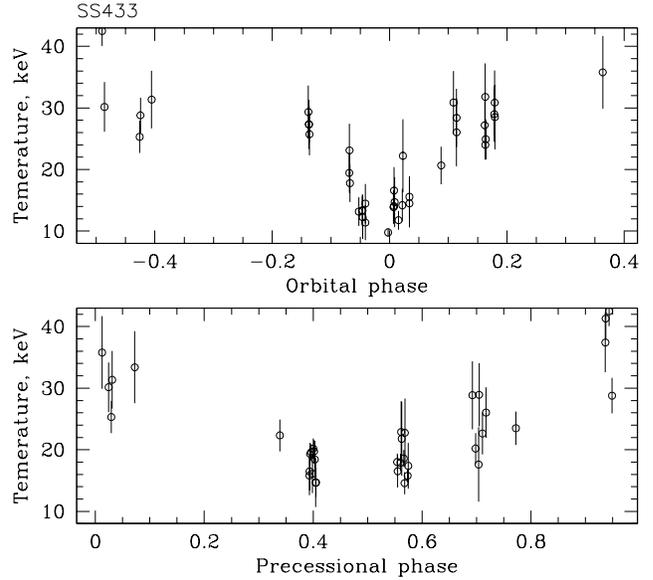}
\caption{The temperature of optically thin thermal plasma
(bremsstrahlung fit to the 10-25 keV RXTE/PCA data)
emission observed from SS433 as a function of orbital and precessional phases.
} \label{temp_prec}
\end{figure}

\begin{figure}[htb]
\includegraphics[width=\columnwidth,bb=33 60 520 450,clip]{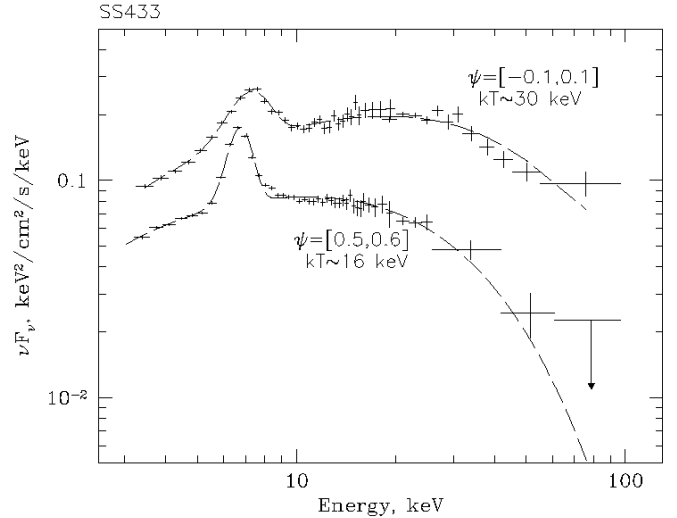}
\caption{Typical spectra of SS433 taken at 
two different disk precession
phases. Only off-eclipse orbital phases $\phi=[0.3-0.7]$ are
selected for this plot. The dashed curves show the fit by the 
bremsstrahlung model  with broad Gaussian line profile. 
The best-fit values of the plasma temperature are quoted.} \label{spectra}
\end{figure}

Orbital variations are not strong, except for the eclipse. In contrast,
the precessional variations are much more pronounced. In
order to demonstrate this, we plot the temperature of
the X-ray emission in Fig. \ref{temp_prec} as a function of precessional and orbital phases.
As the X-ray emission of SS433 is essentially
multi-temperature, we considered  only the high-energy part of the
spectrum (10-25 keV) that probes the hottest, innermost
regions of the X-ray jet. The presented temperatures are best-fit 
parameters of the bremsstrahlung model describing the
observed spectra of SS433 in energy band 10-25 keV. 
Broad-band X-ray spectra of SS433 at different precessional phases
are shown in Fig.\ref{spectra}. To
demonstrate orbital variations, we have chosen the
interval of precessional phases $\psi=[-0.06,0.06]$ (the maximum disk
opening phase). For precessional variations, only
orbital phases $\phi=[0.3,0.7]$ (the off-eclipse phases) have been selected.

It is seen that the best-fit temperature does not exhibit any
correlation with the orbital phase except the X-ray
eclipse (from $\phi\sim -0.15$ to $\phi\sim 0.15$),
while strongly correlating with the precessional phase.

Note that, even at very close both precessional and orbital phases, there
are statistically significant differences in the maximum jet
temperatures. These temperature variations might be caused either by the
nodding motion of the geometrically thick accretion disk or by
chaotic (red noise) variations in the physical parameters of the plasma in the 
jet, similar to
red-noise variations observed in the integrated X-ray flux \citep{mikej05}.

\section{The inner jet tomography by orbital and precessional eclipses}
\label{tomography}

 The precessional motion of the  geometrically thick accretion disk causes partial obscuration 
of the jets.
The upper jet (see Fig.\ref{scheme}) is least eclipsed near
the precessional phase $\psi=0.0$
(i.e. at the maximum disk opening when the upper jet points towards the observer)
and is most eclipsed at $\psi=0.5$
(i.e. when the upper jet points away from the observer). 
Between precessional
phases $\psi\sim0.33$ and $\psi\sim0.66$ (the disk ``edge-on'' or
``crossover'' phases), it is the opposite (lower) jet
that dominates the X-ray emission because it is directed
towards the observer at these phases .

The jet eclipses caused by the optical star and/or accretion disk 
leads to temperature variations of the observed X-ray emission.
Thus, by examining differences between spectra taken during eclipses and
off eclipses we can find the spectral contribution from the innermost 
(hottest) regions of the jet.

\begin{figure*}[htb]
\hbox{
\includegraphics[width=\columnwidth,bb=29 184 563 713,clip]{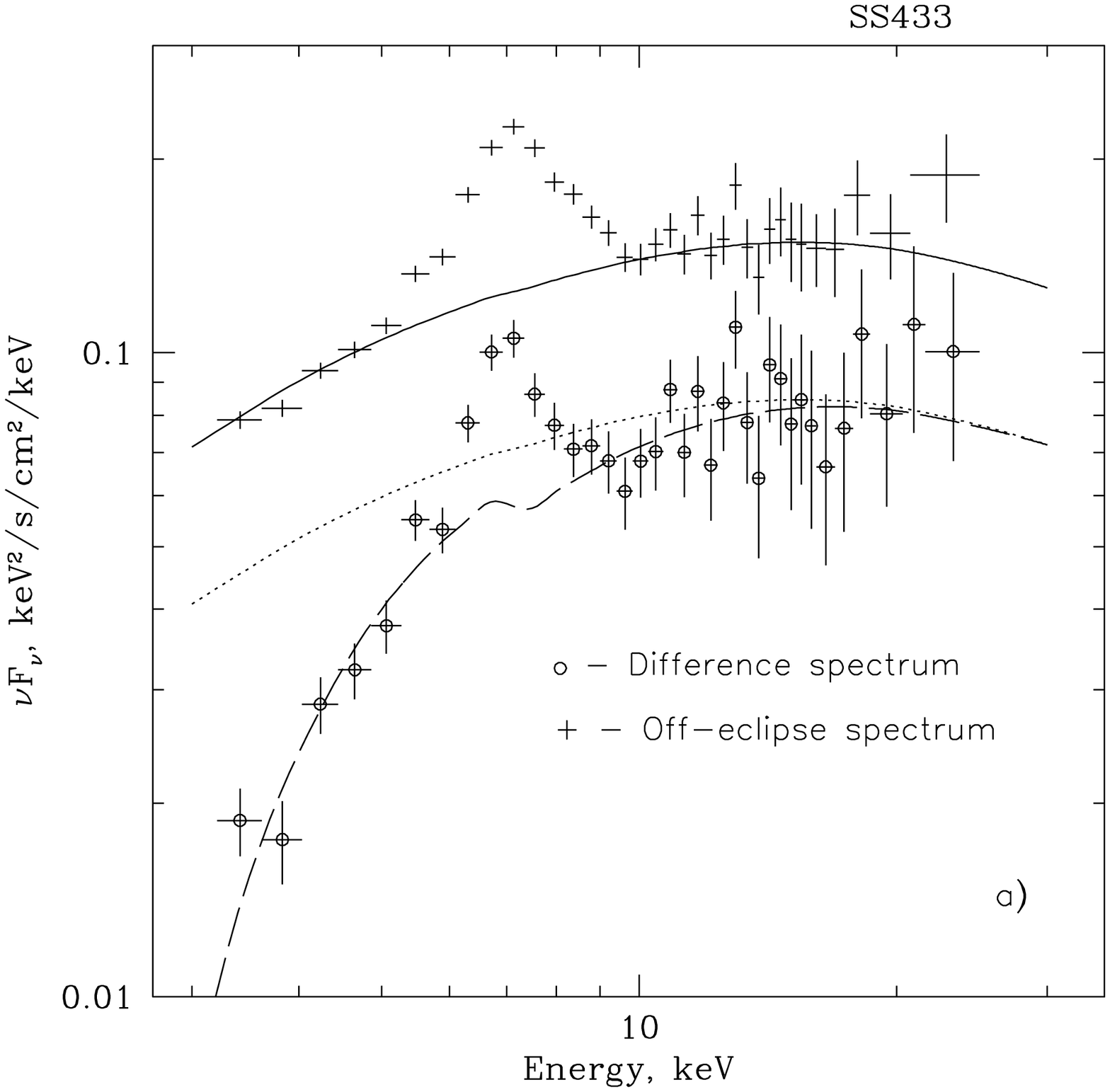}
\includegraphics[width=\columnwidth,bb=29 184 563 713,clip]{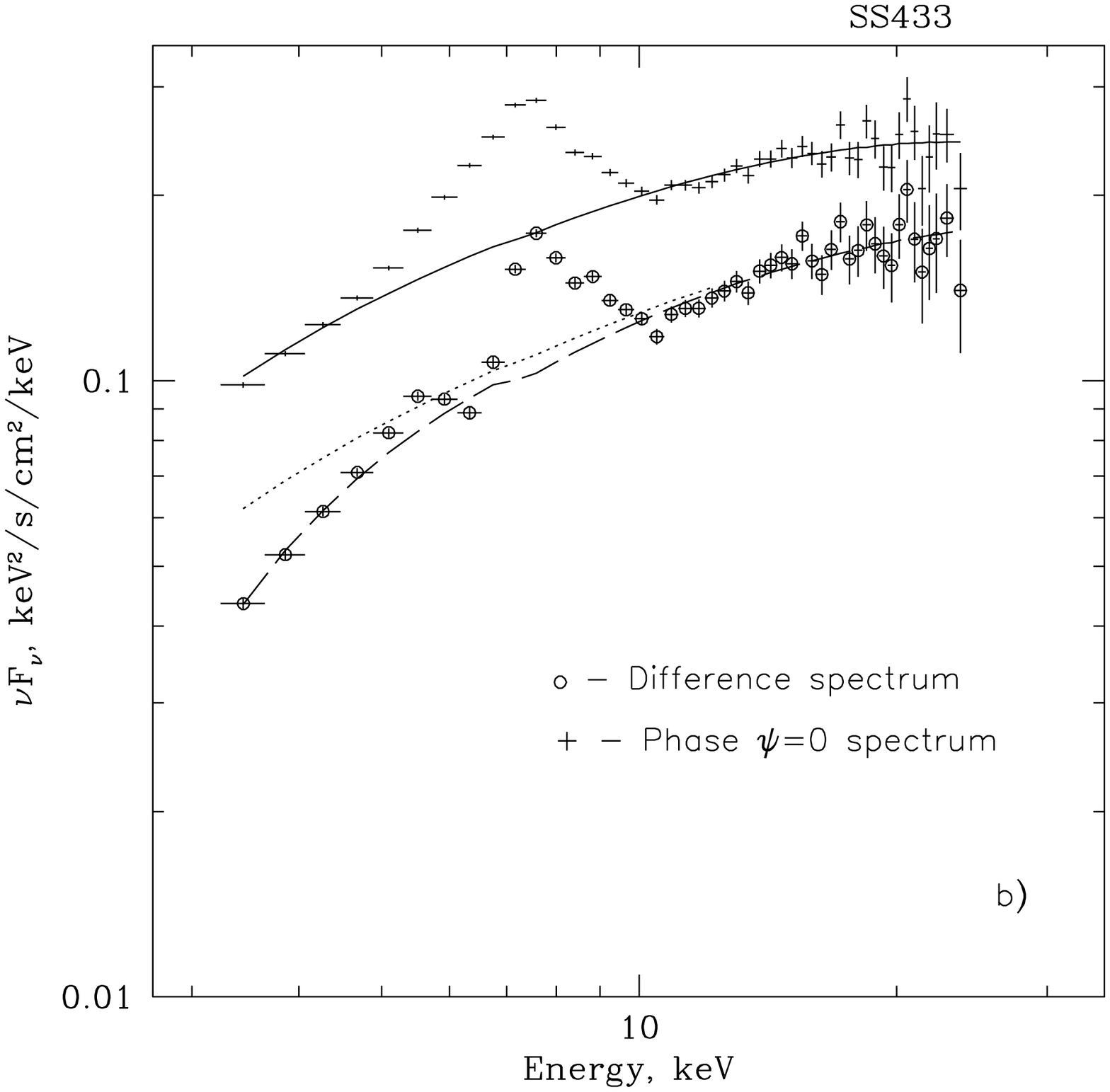}
} \caption{
a) The X-ray spectrum of SS433 immediately after the eclipse
(the orbital phase $\phi=0.114$, crosses) and the spectrum from the inner
part of the jet obtained by subtracting the spectrum taken at $\phi=0.021$
from that at $\phi=0.114$. b) The X-ray spectrum of SS433 at 
precessional phase $\psi\sim0$ and the difference between this spectrum
and that one at $\psi\sim0.4$ (both spectra
were collected over off-eclipse orbital phases). Solid curves show
thermal bremsstrahlung models with photoabsorption fitted to the
data in the 3-5 and 11-25 keV bands. Dotted curves show best-fit
bremsstrahlung models to the ``differential'' spectra with a nominal
photoabsorption column density of $N_{\rm H}=10^{22}$ cm$^{-2}$.}
 \label{dif_spectra}
\end{figure*}

In Fig. \ref{dif_spectra} we present two spectra of the jet in
SS433. In the left panel (a), we plot the difference between spectra
taken during the primary eclipse at $\phi=0.021$ MJD 53581.89 and
immediately after the eclipse at $\phi=0.114$  MJD 53582.94 (open
circles), together with off-eclipse spectrum at $\phi=0.114$
(crosses). The spectra were accumulated over the 1024-s periods.  
Observations were done during the $\psi\sim0$ disk precessional phase.
In the right panel (b), we show the difference between spectra taken at
precessional phases $\psi=0$ (MJD 53076.85) and $\psi=0.4$ (MJD
50878.98)\footnote{These two observations were performed during
different high-voltage epochs of the PCA. We recalculated them to a
single response matrix.  Both spectra were taken off the orbital
eclipses.} together with  spectrum at $\psi=0.114$. The curves 
show best-fit bremsstrahlung plus photoabsorption
models to the data in the 3-5 and 11-25 keV bands where powerful emission lines
do not contribute significantly. The dotted
curves show bremsstrahlung models with a conventional photoabsorption
column density of $N_{\rm H}=10^{22}$ cm$^{-2}$ \cite[e.g][]{kawai89}.

The high energy parts (11-25 keV) of the ``differential'' and time averaged
spectra of SS433 are almost identical because they both are 
determined mainly by emission from the hottest parts of the jet. However, a strong
photoabsorption is observed near the orbital eclipse (the left panel of
Fig.\ref{dif_spectra}). The best-fit value of the absorption column density in
this spectrum is \hbox{$N_{\rm H}=(12.5\pm1.5)\times 10^{22}$ cm$^{-2}$,} much
higher than the conventional value.

The spectrum of the hottest (innermost) part of the jet as derived
from precessional variations off the primary eclipse (the right
panel of Fig.\ref{dif_spectra}) also suggests some photoabsorption, but
with a considerably lowler value -- here the best-fit absorbing column density
is \hbox{$N_{\rm H}=(4.5\pm1.5)\times 10^{22}$ cm$^{-2}$.}

As a result the ``differential'' X-ray spectra of SS433 indicate the presence of
an absorbing material in the SS433 binary system. 
Close to the companion star the density of this
absorbing material is very high, so that
the line of sight close to the stellar surface becomes practically
Compton thick ($N_{\rm H}>10^{23}$ cm$^{-2}$). Note that signatures
of absorbing material near the optical star have
previously been obtained from studies of optical absorption lines
\citep{fabrika97,fabrika97a}. These studies revealed a significant
strengthening in blue-shifted absorption components of P\,Cyg-like line
profiles at orbital phases close to the primary eclipse.

The presence of such a dense absorbing material  near the stellar 
surface can be the signature of
a powerful wind from the optical star.
In order to estimate the possible increase in the eclipsing region
radius with respect to the star's photosphere (in our case with respect to 
the Roche lobe size of the star) due to the donor stellar wind,  
we adopted the wind model of a single $A$ supergiant and took 
parameters of the stellar wind (its mass loss rate and the terminal velocity)  
from the work of \cite{achm97}. From observations and theoretical 
models, it follows that the mass loss rate of an $A$ supergiant strongly 
depends on the mass of the star and its effective temperature and varies from $\sim10^{-9} M_{\sun}/yr$ 
to $\sim10^{-6} M_{\sun}/yr$.
The observed values of the terminal velocity lie in the range 120 - 200 km/s. 
For the velocity law we use  the formula 
\[
{v(r)}={v_{\infty}(1-R_{star}/r)^{\beta}}
\]
where $\beta=0.8$ \citep{achm97}, $v_{\infty}$ is the wind terminal velocity, $R_{star}$ the star radius, 
and $r$ the distance from the companion star. 
From the mass conservation law, we get a formula for the
wind density  $n(r)$:
\[
{n(r)}={{n_0(1-R_{\rm star}/a)^{\beta}}\over{(r/a)^2(1-R_{\rm star}/r)^{\beta}}}
\]
where $n_0$ is the number density at the distance $a$ from the star.
We assume that the line of sight in which the column density of
the stellar wind matter is higher than $N_{\rm H}>10^{24}$ cm$^{-2}$
is opaque for X-rays.
As the spectral type and the mass of the optical star are not known
exactly in the case of SS4333, we consider several cases and obtain the 
following results.
The maximal observed value of mass outflow $\dot{M}=8\times10^{-7}M_{\sun}/yr$ 
(for star type A1 Iae, \citealt{achm97}) results in 
an increase in the X-ray opaque size of the star up to 10 \%. 
The maximal theoretical value $\dot{M}=10^{-6}M_{\sun}/yr$ will lead to  a
20\% increase in the X-ray opaque size of the star.

 Absorption in the inner dense stellar wind material weakens the
assumption that the Roche lobe size of the secondary is a good measure 
of its X-ray opaque size. Note that the X-ray photoabsorption described above should not be
visible in the time-averaged spectrum of SS433 because only emission from the
hottest innermost parts of the jet are screened/absorbed, while outer cooler
parts of the jet, which contribute most to the emission at energies $<3-5$ keV
where the RXTE spectra are most sensitive to absorption, are located much 
farther away from the star and, hence X-ray emission from these
parts is virtually unabsorbed.

\section{Jet eclipses by the thick disk}

Eclipses of the X-ray jet by  a thick disk is a
geometrical effect that depends only on the orientation of the disk
changing with the precessional phase. Therefore,
assuming some reasonable geometry of the thick disk,
observations of SS433 at different precessional phases can be used to
derive emission parameter profiles along the jet.

\subsection{The structure of jets}

We consider the jets in SS433 as conical plasma flows with constant velocity
$v_j=0.26\,c$ along the jet axis. In the direction perpendicular to the jet
axis, matter moves  with a 
constant velocity.
We assume that the jet is uniform in the direction
perpendicular to the jet axis. 

The plasma flow forms a cone
with constant opening angle $2\theta$. The  radius of the jet cross section
is $r=r_{b}+\theta l$, where $r_b$ is the radius of the jet near the
compact object \footnote{This analytical description does not mean
that the jet should start near the compact object.} and $l$  the
distance from the compact object along the jet axis. The value of the
 cone's half-opening angle $\theta=0.61^{\circ}$ is
taken from \cite{marshall02}.

\begin{figure}[htb]
\includegraphics[width=\columnwidth,bb=15 18 570 330,clip]{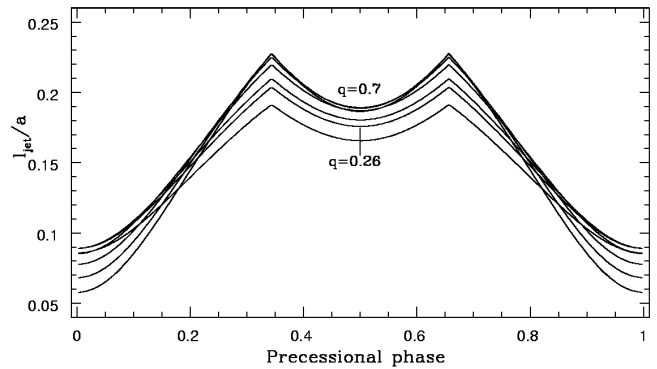}
\caption{The length of the invisible (eclipsed by the disk) part of
the jet as a function of the precessional phase for q=0.2 -- 0.7.} \label{dlina_jet}
\end{figure}

The jet plasma cools due to adiabatic expansion and radiation losses.
When radiation losses become the dominant mechanism of plasma
cooling, thermal instability starts to develop. In this case the 
temperature decreases much more rapidly than in the case of purely adiabatic
expansion \cite[e.g.][]{brinkmann91}. Analysis of the 0.5-10
keV X-ray emission of SS433 with high-energy resolution gratings of \textit{Chandra} suggests that this 
instability does not develop, at least not when the temperature of the 
plasma exceeds $kT\sim 0.5$ keV
\citep{marshall02}. Therefore we assume that the dominant cooling
process in our case (especially in the hottest parts of the jet of
interest here) is adiabatic cooling and we can write
\[
{3\over{2}} {dT\over{T}}={dV\over{V}}=-{2dr\over{r}}\,.
\]

For  adiabatically cooling plasma moving
with a constant jet opening angle,
the plasma temperature $T$ changes with distance from the
compact object $l$ as
\begin{equation}
{T\over{T_0}}={(1+\theta {(l-l_0)\over{r_0}})^{-{4/{3}}}},
\end{equation}
where $T_0$ is the plasma temperature
and $r_0$ the jet cross section radius
at some distance $l_0$ from the central source 
\citep[see also][]{kov89,brinkmann91,kotani96}.

\subsection{Precession of the geometrically thick disk}

The Roche lobe only loosely constrains the
accretion disk radius. For a disk lying in the orbital plane, one
usually adopts Paczynsky's tidal truncation radius as a measure of
the outer disk radius. The accretion disk in SS433 is tilted to the
orbital plane and precesses, so its outer radius can differ from
Paczynsky's estimate. However, according to some numerical models
(e.g. \citealt{nazarenko05}), this difference is not very significant, so
it will be sufficient for our purposes of adopting the disk truncation
radius given by Paczynski (1977):
\[
{{R_{\rm
disk}}\over{a}}=0.112+{0.27\over{(1+q^{-1}})}+{0.239\over{(1+q^{-1})^{2}}}\,.
\]
For a geometrically thick disk ($H/R\sim 1$) and with the understanding that
it should not overfill the Roche lobe of the compact
star, we find
\[
{{H_{\rm disk}}}\la(R_{\rm Roche BH}^2-R_{\rm disk}^2)^{1/2}\,,
\]
where the formula for $R_{\rm Roche BH}$ is adopted from
\citealt{eggleton83}):
\[
{{R_{\rm Roche BH}\over{a}}={{0.49 q^{{2/3}}}\over{0.6
q^{{2/3}}+\ln(1+q^{{1/3}})}}}\,.
\]

In Fig. \ref{dlina_jet} we plot the distance of the
nearest visible jet points from the central source
as a function of the disk
precessional phase for different binary mass ratios $q$.
 The mass ratio is still debatable
on different grounds (see discussion above), so in the present study we
varied the value of $q$ in the range $0.2-0.7$.

Using the formula (1), for given parameters $q,H_{\rm disk}$, and $R_{\rm disk}$,
we can calculate the maximum model temperature of the jet 
visible at different precessional phases. Note that at
precessional phase intervals [0,0.33] and [0.33,0.66] the maximum emission is
provided by different jets (upper or lower). Here and below all
detectable parameters are calculated in the observer's reference
frame, while physical parameters of the jet are given in
the jet's rest frame.

\begin{figure}[htb]
\includegraphics[width=\columnwidth,bb=48 148 565 519,clip]{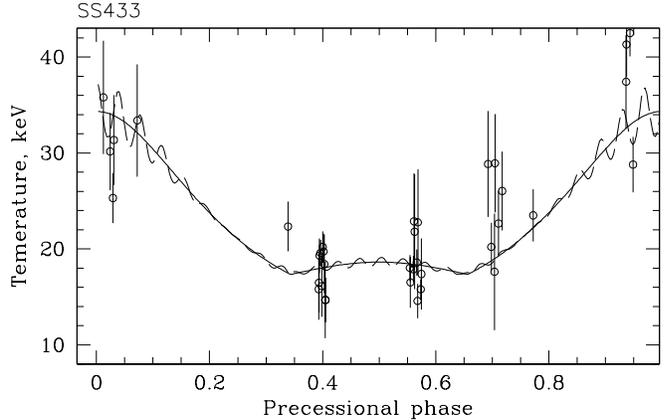}
\caption{The plasma temperature as a function of the precessional phase. The solid curve
shows the model of adiabatic cooling of plasma moving within the cone
with a constant opening angle. The dashed curve shows the same model but
with a nodding motion of thick accretion disk included.}
\label{temp_phase}
\end{figure}

Comparing the profiles of the maximum visible temperatures derived above with
observational data points enables us to find the model 
parameters (see Fig.\ref{temp_prec}).
The best-fit parameters of our model are: $l_0/a=0.06-0.09$ 
(depending on the assumed value of $q$), $T_0=30\pm2$ keV, and 
$r_0/a=(1-1.6)\times10^{-2}$ for the jet opening angle
$\theta=0.61^\circ$. 
The best-fit model is shown in Fig.\ref{temp_phase}.
The limited accuracy of the temperature measurements and incomplete
coverage of the precessional and nutational periods by observations 
preclude us from implementing the disk nodding motion in the
model. For illustrative purposes, in Fig.\ref{temp_phase} we also show the effect of
the nutational variability  by the dashed
curve.

\section{Eclipses by the optical companion}

The X-ray orbital eclipses observed in SS433 are widely used to constrain
the geometry of the binary system
\citep[e.g.][]{stewart87,kawai89}.

Analysis of X-ray eclipses observed at the disk maximum opening (the
precessional phases $\psi\sim0.0$) allows us to get the lower 
limit on the component mass ratio in SS433 independently of the (still
controversial) radial velocity measurements provided by optical
spectroscopy. Indeed, assuming the jet to be infinitely thin in
comparison with the optical star, the duration of the X-ray eclipse
yields only an upper limit on the size of the optical star (however, see the above
 discussion about the likely increase in the X-ray opaque 
size of the secondary star due to its stellar wind).

\begin{figure}[htb]
\includegraphics[width=\columnwidth,bb=48 148 565 620,clip]{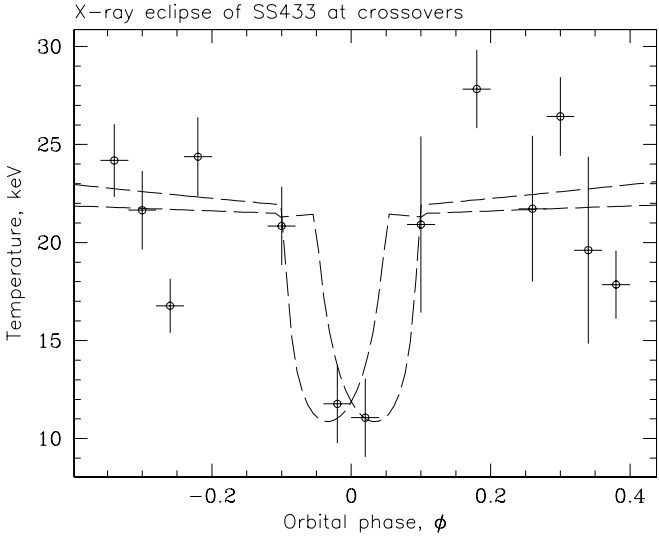}
\caption{Orbital X-ray eclipse of SS433 observed during the
``crossover'' precessional phases $|\psi-0.33|<0.07$ and
$|\psi-0.66|<0.07$. Dashed lines show temperature
profiles obtained in our model.} \label{eclipse_crossover}
\end{figure}

On the other hand, the examination of X-ray eclipses seen at the
disk's ``edge-on'' precessional phases $\psi\sim 0.33$ or $\psi\sim0.66$
(the so-called ``crossover'', because at that time the line of sight 
velocities of both jets are equal and jets lie exactly on the 
plane of the sky) gives us an estimate of the 
star radius in comparison with the accretion disk thickness.
For orbital X-ray eclipses to exist in this precessional phase,
the image of the star on the plane of the sky should not be 
embedded in the image of the thick disk on the plane of the sky. 
Potentially this might give us an upper limit for the value of $q$.

\begin{figure}[htb]
\includegraphics[width=\columnwidth,bb=44 60 565 465,clip]{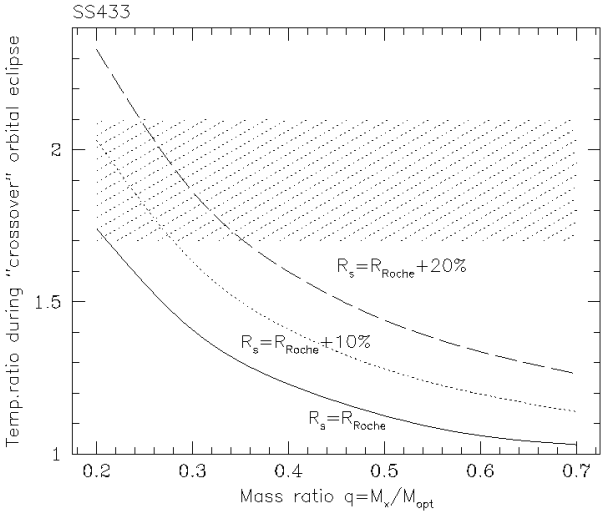}
\caption{The ratio of the maximum visible jet temperatures during the
orbital eclipse in the ``crossover'' precessional phases
$\psi\sim0.33,0.66$ as a
function of the mass ratio $q$. The solid line is obtained
for the star exactly filling its Roche lobe $R=R_{\rm Roche}$, while the
dotted and dashed lines show the model with the $10\%$ and $20\%$
oversized star, respectively. The hatched area shows the observational
constraints on the temperature ratio.}
\label{crossover_temp_ratio}
\end{figure}

\subsection{The ``crossover'' X-ray eclipses}

In Fig.\ref{eclipse_crossover} we present the profile of the maximum
X-ray temperature measured in the orbital eclipse near
precessional phases $\psi\sim0.33$ and $0.66$. Observations with $\Delta
\psi=\pm0.07$ around phases of the exact ``crossover'' were selected.  The
orbital X-ray eclipse is clearly visible and shows an appreciable
depth. The ratio of the maximum jet temperature derived from
off-eclipse observations to that of the eclipsed jet is
$\sim1.9\pm0.2$. Dashed lines in the figure show examples of the
jet maximum visible temperature profiles in the eclipse at precessional
phases $\psi=0.33$ and $\psi=0.66$ obtained in our model. The
temperature profile along the jet was taken as derived in the previous section.
The mass ratio $q=0.2$ was  adopted here for clarity,  and also we assumed that the size 
of the star is 10 \% larger than its Roche lobe. We have not attempted to best-fit
the observed points as they were collected during
observations performed at different nutational and precessional
phases. Note that the model X-ray eclipse profiles during crossovers are not
symmetric. This happens because the jets in these two cases are
inclined differently with respect to the binary orbital plane.

The ratio of the maximum jet temperatures in the eclipse and off the eclipse
is plotted in Fig.\ref{crossover_temp_ratio} as a function of the
binary mass ratio $q$. The solid curve shows the ratio obtained
under the assumption that the size of the star equals the volume-averaged
radius of the Roche lobe. The value of $q$ affects the depth of
the ``crossover'' eclipse via the size (thickness) of the accretion
disk and the secondary star.

We have shown in Sect. \ref{tomography} that the assumption that the Roche 
lobe size of the secondary is a good measure of the X-ray opaque
size of the star might be wrong due to absorption in 
the inner, dense stellar-wind material. Therefore in Fig.
\ref{crossover_temp_ratio} we also
show models of the ratio of maximum jet temperatures in and out of the
eclipse assuming the star radius to be $R=1.1 R_{\rm Roche}$ and
$R=1.2R_{\rm Roche}$ by the dotted and dashed lines correspondingly (see discussion in  
Sect. \ref{tomography}).

From Fig.\ref{crossover_temp_ratio} we can conclude that the
binary mass ratio in SS433 
cannot be directly determined from the depth or duration of the X-ray eclipses.
The result depends on the extent of the eclipsing region over the stellar Roche lobe. 
However, if the binary mass ratio is significantly higher than $q\sim0.3-0.35$,
the eclipsing region size seems to be too big, $R>>1.2R_{\rm Roche}$, which
is not very likely. Comparison of the observed temperature ratio with our model
suggests that the X-ray opaque size of the star should be larger than 
$R/a\ga 0.5-0.55$.

\subsection{X-ray eclipse at $\psi\sim0$}

\begin{figure}[htb]
\includegraphics[width=\columnwidth]{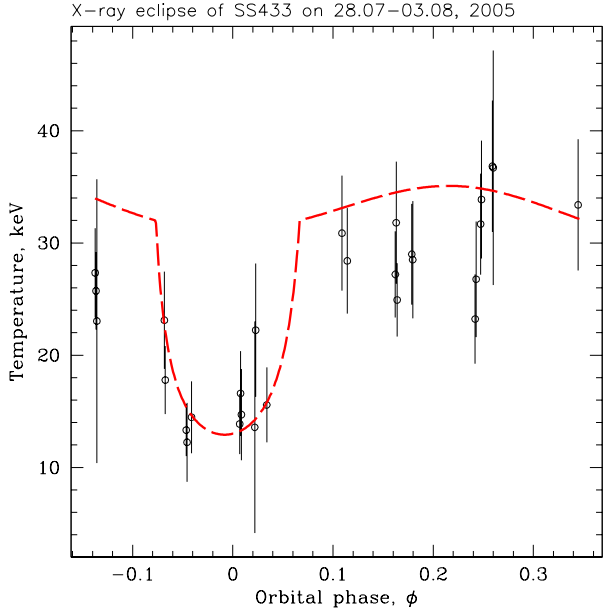}
\caption{The jet eclipse profile by the optical companion for
$q=0.26$. The dashed line shows the best-fit model.} \label{eclipse}
\end{figure}

 In order to obtain a high quality profile of the jet maximum temperature
across a single eclipse, we performed a set of dedicated RXTE
observations of SS433 on July 28 -- Aug. 3, 2005 near the precessional phase
$\psi\sim0$. The temperature profile obtained from these observations is 
presented in Fig.\ref{eclipse}.
Using the temperature distribution along the jet length
$l$ derived in the previous section, we calculated the expected
behavior of the maximum jet temperature during the eclipse for different sizes
of the eclipsing region $R$.

The comparison of the model with obtained data points yielded the 
best-fit value $R/a=0.53\pm0.02$ ($\chi^2=34$ for 24
degrees of freedom) for $q=0.26$, which is $\sim$7\% higher than the
size of the Roche lobe of the secondary would have at such $q$. The best-fit
model is shown in Fig.\ref{eclipse}. For demonstration purposes we
included the effect of accretion disk nodding motion in the model. 

It is interesting to note that the quality of the fit can be
improved if we assume {\bf a} non-spherical shape of the eclipsing region (star
plus inner wind). This is a plausible geometry if the wind
density from equatorial regions of the star is higher than from
polar regions.

\section{Conclusion}

X-ray emission of SS433 demonstrates both chaotic (aperiodic) and
periodic variations. In this paper, we used extensive observational
data obtained by RXTE to studing systematic variations in the X-ray
emission of SS433 caused by precessional and orbital motions in the
system.
\begin{itemize}
\item
By comparing X-ray spectra of SS433 near and in the eclipse, we
obtained strong signatures of a significant ($N_H>10^{23}$
cm$^{-2}$) photoabsorption of X-rays near the companion star. We
argue that this might be caused by the presence of a dense stellar wind
from the companion star. The mass loss rate in the
wind needed to explain the observed photoabsortion is
$\dot{M}\sim 10^{-6}$ M$_{\odot}$/yr.

\item 
 From detection of strong photoabsorption near the secondary star,
we conclude that the X-ray opaque size of the secondary might be
significantly larger than the Roche lobe size of the star. 
Therefore one should be cautious when estimating the Roche
lobe size of the secondary from X-ray orbital eclipses.

\item By both assuming the shape of the geometrically-thick accretion
disk restricted by the Roche
lobe size of the compact star and using the RXTE
observations of SS433 at different precessional phases, we recovered
the temperature profile of plasma along the jet. 
The obtained maximum temperature visible in the jet is $T\sim30$ keV at a
distance of $l/a=0.06-0.09$ from the compact object. The radius of the jet at
this distance is $r_0/a=0.01-0.016$. Here the jet was assumed to have 
a constant opening angle.

\item
We reliably detected orbital X-ray eclipses in SS433 during 
the ``crossover'' precessional phases when the X-ray
jets and the axis of the accretion disk lie exactly on the plane of
the sky. The observed depth of the X-ray eclipse combined with
our assumptions about the accretion disk thickness implies that the
size of the star is larger than $R/a\ga0.5$, yielding an upper bound
on the mass ratio of the components in SS433 $q<0.3-0.35$, assuming
that the radius of the eclipsing region (star plus inner wind) cannot be
much larger than $1.2R_{\rm Roche, secondary}$.

\end{itemize}

\begin{acknowledgements}
Authors are grateful to the anonymous referee for very valuable remarks, 
which helped to improve the manuscript.
This research made use of data obtained from High Energy Astrophysics
Science Archive Research Center Online Service, provided by the
NASA/Goddard Space Flight Center. This work was supported by
RFBR grants N\,04-02-16349, N\,05-02-19710, N\, 06-02-16025, N\, 04-02-17276, and N\,
04-02-16720, and by joint RFBR/JSPS grant N\,05-02-12710.

\end{acknowledgements}

\end{document}